\documentclass[runningheads,citeauthoryear]{apinv}
\usepackage{epsfig,cite,graphics}
\usepackage{marvosym} 
\usepackage[utf8]{inputenc}

\begin{document}
	
	\title{Chi-square test of the relativistic precession model  through the neutron star IGR~J17511-3057}
	\titlerunning{Chi-square test of the RPM}
	\author{Ivan Z. Stefanov}
	\authorrunning{Ivan Z. Stefanov}
	\tocauthor{Ivan Z. Stefanov} 
	\institute{Deparment of Applied Physics, Faculty of Applied Mathematics and Informatics,\\ Technical university of Sofia,\\ 8, Snt Kliment Ohridski Blvd, Sofia 1000, Bulgaria   \newline
		\email{izhivkov@tu-sofia.bg}    }
	\papertype{Submitted on xx.xx.xxxx; Accepted on xx.xx.xxxx}	
	\maketitle
	
	\begin{abstract}
		The aim of the current paper is to apply Bambi's  method (Bambi, 2015) to a source which contains two or more simultaneous triads of variability components.  The joint $\chi^{2}$ variable that can be composed in this case, unlike some previous studies, allows the goodness of the fit to be tested. It appears that a good fit  requires one of the observation groups to be disregarded. Even then, the model's prediction for the mass of the neutron star in the accreting millisecond pulsar IGR J17511-3057 is way  too high to be accepted.
	\end{abstract}
	
	\section*{1. Introduction}
	The presence of a simultaneous triad of quasi-periodic oscillations (QPOs) consisting of a pair of twin high frequency QPOs (HF QPOs), a upper and a lower one, and one low frequency QPO (LF QPO) in the spectra of low-mass X-ray binaries gives us the opportunity to measure with great precision the spin and mass of their central compact object  (a black hole or a neutron star)  as it was shown by Motta et al. (2014). They applied the relativistic precession model to the simultaneous triad present in the X-ray power-density spectrum of the black hole binary GRO J1655-40. As a confirmation of the adequacy of the method and the model the authors compare their estimation of the mass of the studied black hole to an earlier independent measurement and obtain very good agreement. 
	
	Soon after the work of Motta et al.(2014), Bambi (2015) studied the same stellar-mass black hole. He exploited the same simultaneous triad of QPOs and used the same model but proposed a different method, much more efficient from a numerical point of view, for calculation of the uncertainty of the estimated parameters -- mass and spin. Bambi (2015) proposed a merit function which allows confidence level regions to be plotted in the joint mass-spin parameter space. It proves very useful for the estimation of the uncertainty of the  parameters but it has zero degrees of freedom (dof) and cannot be treated as a standard $\chi^{2}$ variable, so it does not allow us to apply a statistical test for the adequacy of the model.   
	
	The presence of simultaneous twin HF QPOs in the X-ray spectrum of BHBs is a rather rare event ( Belloni, 2012; Ingram \& Motta, 2019). Simultaneous triads are even less frequent. GRO J1655-40 is, as far as we know, the only BH which has a confirmed simultaneous triad of narrow timing features. (This is why the QPOs of GRO J1655-40 are so vastly exploited.) Hence, it appears very unlikely to find a BHB with more than on simultaneous triad of QPOs in its X-ray power density spectrum.
	
	One could increase the degrees of freedom of the merit function by the inclusion of data from other, though incomplete, triads of QPOs. Such an incomplete triad is present in the spectrum of GRO J1655-40 (Motta, 2014). It consists of a upper HF QPO and a C-type LF QPO. This pair has been added in the definition of a merit function, again with zero degrees of freedom,  in (Bambi \& Nampalliwar, 2016). It increases the number of terms by two but also requires the inclusion of one more parameter --  radius of the orbit at which it originates, which means that due to the inclusion of this pair the degrees of freedom rise by one. In this situation the data allows one more free parameter to be added in the analysis, and hence deviations from the Kerr metric to be tested (Chen et al., 2016; Jiang, 2021). Another possibility to increase the number of studied free parameters is to supplement the simultaneous triad of QPOs with data from other measurements, such as the mass, as it was done by Bambi, (2015) and Allahyari et al. (2021). The authors of the latter paper use Bayesian approach and study the posterior probability distribution of the estimated parameters. 
	
	As it was recently shown, in (Ghasemi-Nodeli et al., 2021) the method of Bambi can be  generalized to the case in which more than one triad of simultaneous QPOs is observed. In this paper, however, the datasets which contain several triads were not taken from observations but were rather synthesized.
	
	Taking into account that the RPM can  explain also the kHz QPOs, the analogue of the black-hole HF QPOs,  and the low-frequency variability components of neutron stars we search for simultaneous triads among them. For recent observational interest in X-ray binaries see (Nikolov, 2020) and references therein. One possible candidate is the accreting millisecond X-ray pulsar  IGR J17511-3057. It contains three triads of supposedly simultaneous variability components. Two of the frequencies in those triads could be classified as twin kHz QPOs, according to one of the interpretations given by (Kalamkar et al., 2011). The third component in the groups of simultaneous observations of this source is not the C-type QPO that usually completes the triads described by the RPM, however, but a ``hump'' $L_h$ variability component.   
	
	The aim of the current study is to apply the method presented in (Bambi, 2015) to obtain confidence level plots in the spin-mass parameter space for IGR~J17511-3057. We would like to answer several questions. First, are the estimates for the mass and the spin of this object that come from the different triads in agreement with each-other? 
	
	The $\chi^{2}$ variable obtained from the combination of two triads of frequencies contains six terms. In order to describe them we need at least four parameters (in the case of Kerr metric) -- the mass and the spin of the central object, and the radii of the two orbits on which the observed triads originate. Hence, the joint $\chi^{2}$ variable has two degrees of freedom. If we combine all three of the simultaneous triads, the results is a $\chi^{2}$ variable with four degrees of freedom.  In these cases, we have the opportunity to apply the the $\chi^{2}$ goodness-of-fit test. So, the second question that we ask is whether the minimum value $\chi^2_{\rm min}$ allows us to accept the model?
	
	There is one more condition which a feasible model must satisfy -- to give a reasonable estimate for the mass of the neutron star, not exceeding the commonly accepted upper bound of three Solar masses. 
	
	We should clarify that in the current study black hole binaries and neutron star binaries are treated on equal footing in the sense that for the central object same metric is used -- Kerr, and the same model is applied for the description of the observed QPOs, namely the relativistic precession model (RPM). As it was shown in, (Stuchl\'{\i}k \& Kolo\v{s}, 2015), the former assumption is justified for the study of QPOs of black holes with spin lower than 0.4.
	
	The paper is organized as follows. The accreting millisecond pulsar IGR J17511-3057 and its narrow timing features  are briefly presented in Section 2.  A brief presentation of the RPM follows in Section 3. Section 4 is dedicated to the estimates of the spin relation coming from the different observations and to the role of uncertainties.  Then come  the Discussion and the Conclusion.
	
	In this paper all the masses are scaled with the Solar mass, the radii are scaled with the gravitational radius $r_{\rm g}\equiv G M/c^2$, the specific angular momentum $a\equiv J/c M^2$ and measure units in which $G=1=c$ are used.
	\section*{2. IGR J17511-3057}\label{IGR}
	
	The X-ray timing of the millisecond accreting pulsar IGR J17511-3057 has been studied by Kalamkar et al. (2011). They present the results from 71 pointed observations of this object with the RXTE PCA. For the sake of better statistics close in time and color (We refer the reader to (Ingram \& Motta, 2019) for more details on the color and the classification of the states of X-ray sources.) observations have been combined into seven groups of 7 to 15. Following (Belloni, Psaltis, \& van der Klis (2002)) the authors fit the  power spectrum of each group by a multi-Lorentzian function -- a sum of several Lorentzians. For the particular groups of observations considered here, namely 1, 2 and 7, the number of Lorentzians is five. Each Lorentzian corresponds to a different component in the power density spectrum (PDS). The different components have been identified with the help of correlation diagrams (Wijnands \& van der Klis 1999; Psaltis et al. 1999; van Straaten et al. 2005) (See Section 3.3 in (Kalamkar et al. 2011) for more details.).

	In the current paper we are searching for simultaneous triads of characteristic frequencies, namely a pair of twin kHz QPO (which are the analogue of the black-hole HF QPOs for neutron stars) and one more low-frequency feature such as a low-frequency QPO or, as in this case, a "hump" feature $L_h$. Such components have been seen in three of the seven groups of observations - 1, 2 and 7. The presence of such triads would allow one to apply the relativistic precession model to describe them and, hence, to obtain estimates for the model parameters -- the mass and the spin of the black hole and the radius of the orbit of origin of the observed features, resulting from the motion of inhomogeneities in the disk, as the model states.
	When it comes to the low-frequency sector the relativistic precession model has been applied for the explanation of the C-type low-frequency QPOs, mostly, but also of the HBOs (horizontal branch oscillations) of Z sources (Stella \& Vietri, 1998; Stella et al., 1999; Stella \& Possenti, 2009) and $L_h$ of the atoll source 4U 1608–52 (van Straaten et al., 2003). 
	
	The studied source IGR J17511-3057 has no LF QPOs in groups 1, 2 and 7. We have chosen them in particular due to the presence of twin kHz QPOs. The only possibility that remains in this case, is to complement the desired triads with the hump features $L_h$ that are present in these groups.

	The values of the characteristic frequencies and their uncertainties are reproduced here in Table \ref{tab}.
	\begin{table}
		\centering
		\caption{Features in the PDS of IGR J17511-3057 }
		\label{tab}
		\begin{tabular}{ccc}
			\hline
			group & $\nu_{\rm max}$, (Hz)  & Identification \\
			\hline
			1& $251.8 \pm 13.9$ & $L_u$   \\
			& $139.7 \pm 4.2$  & $L_l$   \\
			& $6.4 \pm 0.6$  & $L_h$   \\
			\hline 
			2& $272.2 \pm 13.9$ & $L_u$   \\
			& $129.9 \pm 11.0$ & $L_l$           \\
			& $5.2 \pm 0.2$  & $L_h$   \\
			\hline 
			7& $179.9 \pm 14.9$ & $L_u$          \\
			& $72.5 \pm 4.9$&    $L_u$           \\
			& $13.7 \pm 2.5$  & $L_h$   \\
			\hline
		\end{tabular}
	\end{table}
	As Kalamkar et al. (2011) state the interpretation of the highest frequency QPOs is not certain. One possibility is that these pairs represent twin kHz QPOs. It is possible, however, that the second highest frequency QPOs are hecto Hz QPOs. There is also one more scenario. According to Scenario 4 in Kalamkar et al. (2011) the two highest frequency QPOs in groups 1 and 2 are hecto Hz QPOs, while the two highest frequencies of group 7 cannot be identified.  
	
	We have to admit also that, according to Kalamkar et al. (2011), the identification of the $L_h$ features is also not certain.
	\section*{3. Relativistic precession model}\label{RP}
	
	Stella \& Vietri proposed the RPM (Stella \& Vietri, 1998; Stella et al., 1999; Stella \& Possenti, 2009) in their attempt to explain the correlation between the low frequency QPOs and the lower kHz QPOs that has been reported by Psaltis, Belloni \& van der Klis (1999; van der Klis 2006)  for a large number of neutron-star sources. 
	Soon after that they adapted it to black holes (Merloni et al., 1999). 
	
	According to the RPM, which belongs to the group of the hot spot models, the X-ray flux of black holes and neutron stars is modulated by the motion inhomogeneities in the inner region of the accretion disk where most of the energy is released. The frequencies that appear in the power density spectra are related to the geodesic orbital, epicyclic and precession frequencies. 
	
	We could say that the RPM is a collection of three models -- one for the low-frequency narrow timing components such as a LF QPO, a horizontal branch oscillation or a hump component $\nu_{\rm h}$, and two more -- for the lower $\nu_{\rm l}$ and the upper $\nu_{\rm u}$ twin kHz QPOs. These frequencies are attributed to the nodal precession $\nu_{\rm nod}=|\nu_{\rm \phi}-\nu_{\rm \theta}|$, the periastron precession $\nu_{\rm per}=\nu_{\rm \phi}-\nu_{\rm r}$ and the orbital $\nu_{\rm \phi}$ frequencies, respectively.

	The radial and the vertical epicyclic frequencies, $\nu_{\rm r}$ and $\nu_{\rm \theta}$, have been obtained for the Kerr space-time in (Aliev \& Gal'tsov 1981; Aliev et al., 1986; Aliev et al., 2013). They are reproduced here in the Appendix.
	\section*{4. Modeling of data}\label{Modeling}
	\subsection*{4.1 Each triad separately}
	In this section we follow the idea presented in (Bambi, 2015) and define the following merit function for the $\rm i$-th triad of QPOs, where the index "$\rm i$" designates the groups of observations which contain simultaneous triads, namely 1st, 2nd and 7th,  
	\begin{eqnarray}\label{merit_Bambi}
		\chi_{\rm i}^2(a, M, r_{\rm i})&=&\frac{\left(\nu_{\rm h}-\nu_{\rm h,i}^{\rm obs}\right)^2}{\sigma_{\rm h}^2}+\\ \nonumber &+&\frac{\left(\nu_{\rm l}-\nu_{\rm l,i}^{\rm obs}\right)^2}{\sigma_{\rm l}^2} +\frac{\left(\nu_{\rm u}-\nu_{\rm u,i}^{\rm obs}\right)^2}{\sigma_{\rm u}^2}. 
	\end{eqnarray}
	Here the observed frequency of the hump feature is designated as  $\nu_{\rm h,i}^{\rm obs}$, while those of the two kHz QPOs are, repectively, $\nu_{\rm l,i}^{\rm obs}$ and $\nu_{\rm u,i}^{\rm obs}$. The model frequencies are  $\nu_{\rm h}$,  $\nu_{\rm l}$ and $\nu_{\rm u}$, respectively. 
	
	The merit function (\ref{merit_Bambi}) contains three terms -- one for each of the frequencies in the triad. The optimal values of the parameters $M$, $a$ and the radius $r_{\rm i}$ of the orbit on which the $\rm i$-th pair originates  are those which minimize (\ref{merit_Bambi}). The number of optimized parameters is equal to the number of terms, so the merit function has zero degrees of freedom in this case. It is not an authentic $\chi^{2}$ variable. 
	
	As it was demonstrated in (Bambi, 2015) we can define a $\chi^{2}$ variable (with two rather than three degrees of freedom) which can be used for the obtaining of confidence level regions in the plane of the parameters $a$ and $M$ in the following way  
	\begin{equation}\label{Delta_chi}
		\Delta \chi_{\rm i}^2\equiv\chi_{\rm i}^2-\chi_{\rm i, min}^2.
	\end{equation}
	Here $\chi_{\rm i, min}^2$ is the minimum value of (\ref{merit_Bambi}), which is actually zero whenever the system defined by the three frequencies and the three observed values has a solution, while $\chi_{\rm i}^2$ is the value of (\ref{merit_Bambi}) obtained for given values of $M$ and $a$ and minimized with respect to  $r_{\rm i}$. (The geometrical interpretation of this condition is briefly commented in (Stefanov, 2020a).) The $\chi^{2}$ variable (\ref{Delta_chi}) has three terms and depends on three parameters. Once again, two of the parameters, $M$ and $a$ can vary freely, but the value of $r_{\rm i}$ is chosen so as to minimize $\Delta \chi_{\rm i}^2$. The minimization of $\Delta \chi_{\rm i}^2$ with respect to one of the parameters reduces the degrees of freedom by one, or ${\rm dof}=3-1=2$. 
	For more details\footnote{Prof. Peter Scott has a document entitled ``\textit{Physics 133 Lab manual}'' on his personal webpage \texttt{\small http://scott.physics.ucsc.edu/} which is an excellent pedagogical text on the chi-square goodness-of-fit test.} on the $\chi^{2}$ goodness-of-fit test we refer the reader to (Press et al., 2007), (Bevington \& Robinson, 2003) and (Stefanov \& Tasheva, 2022). 
	
	In 68.3\% of the cases the $\chi^{2}$ variable $\Delta \chi_{\rm i}^2$ takes values lower than or equal to 2.3. This confidence level, corresponding to 1-$\sigma$, defines a region in the space of parameters $M$ and $a$. Its boundary is represented by an ellipse on Figure 1 for each one of the three groups of observations -- 1, 2 and 7.
	\begin{figure}[ht!]
		\begin{center}
			\centering{\epsfig{file=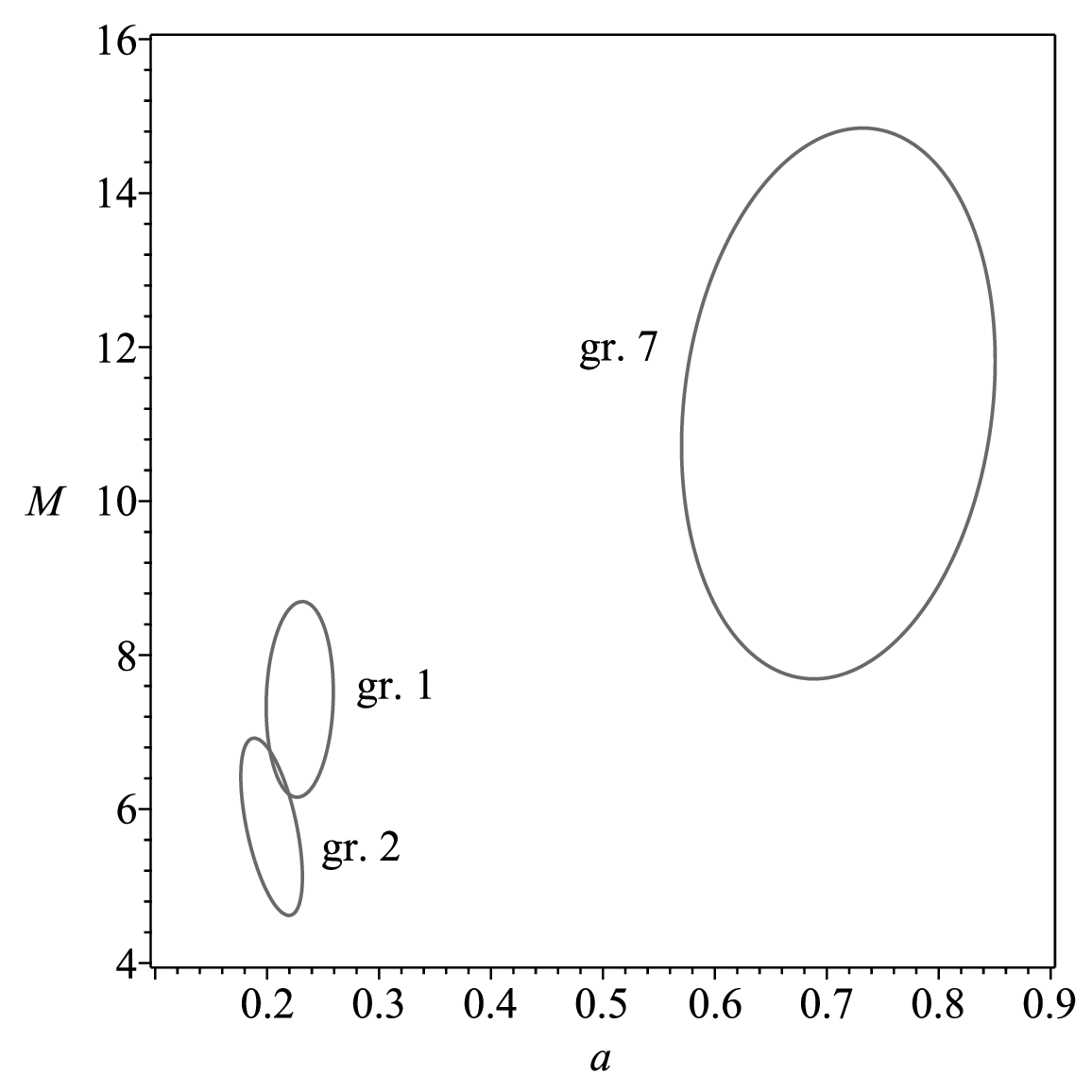, width=0.65\textwidth}
				\caption{The 1-$\sigma$ covariance plots obtained with the data from the triads from groups 1, 2  and 7.}}\label{1triad}
		\end{center}
	\end{figure}
	Here and bellow we choose to present only the 1-$\sigma$ confidence regions and not to include the 2-$\sigma$ and 3-$\sigma$ ones, as customary, in order to avoid cumbersome figures. Besides, in this work we are searching for conflicts between the predictions coming from the different measurements which are more likely to manifest themselves when more conservative constraints on the parameters are used. For the production of the confidence regions (the error ellipses) we follow the procedure from the prominent book "Numerical Recipies" (Press et al., 2007) in which the $\chi^{2}$ variable is expanded to the leading, actually second, order terms. As a result the confidence regions are ellipses. This method has been follow also by (Maselli, 2017), for example. Bambi (2015) and most of the works that cite him choose a different approach and plot the``full'' 2D confidence level regions which have irregular shape. This approach can be seen also in Kyurkchieva et al. (2020) where constraints for the  possible mass ratios and the inclinations of two short-period eclipsing stars are given.
	
	As it can be seen from Figure 1  the confidence level regions coming from the different groups of observations, which can be treated as three independent experiments or measurements, do not overlap. Groups 1 and 2 are in partial agreement but group 7 appears to be an outlier.
	
	The constraints on $a$ resulting from groups 1 and 2 are moderate and comply with the requirement $a\leq0.4$, which guarantees the validity of the Kerr metric for the evaluation of the epicyclic frequencies of a neutron star. Group 7, however, predicts much higher values for the spin, which give us reason to question the applicability of Kerr metric in this case. For all of the three measurements the constraints on $M$ are way too high for a neutron star. Even the stiffest equations of state predict masses lower than 3 Solar masses.
	\subsection*{4.2 Combinations of three triads}
	In this subsection we take advantage of the fact that the X-ray spectrum of IGR J17511-3057 contains several triads of QPOs which can be combined. The joint merit function that results from the combination of the three pairs of independent observations this time, unlike the previous case, is an authentic $\chi^{2}$ variable with four degrees of freedom, a fact which gives us the opportunity to test the validity of the model. The combination of all three triads of QPOs allows us to define the following  $\chi^{2}$ variable 
	\begin{equation} \label{merit_three}
		\chi^2_{\rm three}(a, M, r_{1}, r_{2}, r_{7})=\sum_{\rm i = 1,2,7}\chi_{\rm i}^2,
	\end{equation}
	where the $\chi_{\rm i}^2$-s are given by equation (\ref{merit_Bambi}).
	It is a $\chi^{2}$ variable  with dof=4, since it has nine terms and five parameter --  $a$, $M$ and the radii of origin of the three triads in the combination $r_1$, $r_2$ and $r_7$ which are optimized.
	The minimum value of $\chi^2_{\rm three}$ (\ref{merit_three}) is
	\begin{eqnarray}
		\chi_{\rm three, min}^2=22.9. \label{chi3min}
	\end{eqnarray}
	The quality of the fit appears to be poor. The critical value for a $\chi^{2}$ variable with dof=4, at 5\% level of confidence, is 9.5, while the minimum value (\ref{chi3min}) that we obtain,   22.9, is significantly higher than that. 
	
	Confidence limits of the estimated values of $a$ and $M$ can be obtained, just as in the previous subsection, with the help of the $\chi^{2}$ variable
	\begin{equation}\label{Delta_chi_2_three}
		\Delta \chi_{\rm three}^2\equiv\chi_{\rm three}^2-\chi_{\rm three, min}^2.
	\end{equation}
	The error ellipse in the $a-M$ parameter space obtained with (\ref{Delta_chi_2_three}) is given on Figure 2 with the thick solid line.
	
	As it can be seen on Figure 2 the estimation for the mass, given by the projection of the error ellipse on the vertical axis,  is again too high for a neutron star. The spin is moderate, lower than 0.4, and justifies the application of the Kerr metric. 
	
	Interestingly, the thick solid error ellipse is pulled to the left, towards the moderate values of the spin, even though group 7 is also taken into account. The reason for this are the significantly smaller uncertainties and, hence, greater weights of the hump components in groups 1 and 2, which favor moderate spin, in comparison to that of group 7, which requires high spin.

	\subsection*{4.3 Combinations of two triads}
	The reason for the poor fit from the previous subsection might be the presence of an oultier, such as group 7, for example. In order to test this possibility we remove one of the groups and define another merit function by the combination of two different groups of observations   
	\begin{equation} \label{merit_two}
		\chi^2_{\rm two}(a, M, r_{\rm i}, r_{\rm j})=\chi_{\rm i}^2+\chi_{\rm j}^2,
	\end{equation}
	where, the $\chi_{\rm i}^2$-s are given by equation (\ref{merit_Bambi}), ${\rm i}, {\rm j} = 1, 2, 7$ and ${\rm i} \not= {\rm j}$. We take into account all possible combinations of two groups of observations that can be composed, namely 1 and 2, 1 and 7 and 2 and 7, and obtain three $\chi^{2}$ variables $\chi^2_{\rm two}$. Each of them has six terms and is optimized with respect to four parameters -- $a$, $M$ and the radii of origin of the two triads in the combination, i.e. ${\rm dof} = 6-4=2$.
	
	The minimum values of the $\chi^{2}$ variable obtained by the combination of groups 1 and 2, 1 and 7 and 2 and 7, respectively are:  
	\begin{eqnarray}
		\chi_{\rm two, 12, min}^2=3.91, \label{chi12min} \\
		\chi_{\rm two, 17, min}^2=17.1, \label{chi17min} \\
		\chi_{\rm two, 27, min}^2=20.2. \label{chi27min}
	\end{eqnarray}
	For two degrees of freedom the critical value of the $\chi^{2}$ variable corresponding to 5\% confidence level is 5.991. As it can be seen form eqs. (\ref{chi12min})--(\ref{chi27min}) the inclusion of groups 7 in the combination yields values for $\chi^2_{\rm min}$ which are higher than the critical one. In other words, groups 1 and 2 can be reconciled (See the value of \ref{chi12min}.), but, as it appears from (\ref{chi17min}) and (\ref{chi27min}), 7 is not in agreement with them.

	Just as in the previous cases, for all of the  combinations we obtain 1-$\sigma$ confidence ellipses  in the $a-M$ space through level plots of the following function  
	\begin{equation}\label{Delta_chi_2_two}
		\Delta \chi_{\rm two}^2\equiv\chi_{\rm two}^2-\chi_{\rm two, min}^2,
	\end{equation}
	expanded in Taylor series up to second order terms in $a$ and $M$.
	The error ellipses are given again on Figure 2. On this figure the ellipse obtained with groups 1 and 2 is represented by a dash-dotted line, with 1 and 7 by a dashed line and with the last combination -- 2 and 7, by a dotted line. 
	\begin{figure}[ht!]
		\begin{center}
			\centering{\epsfig{file=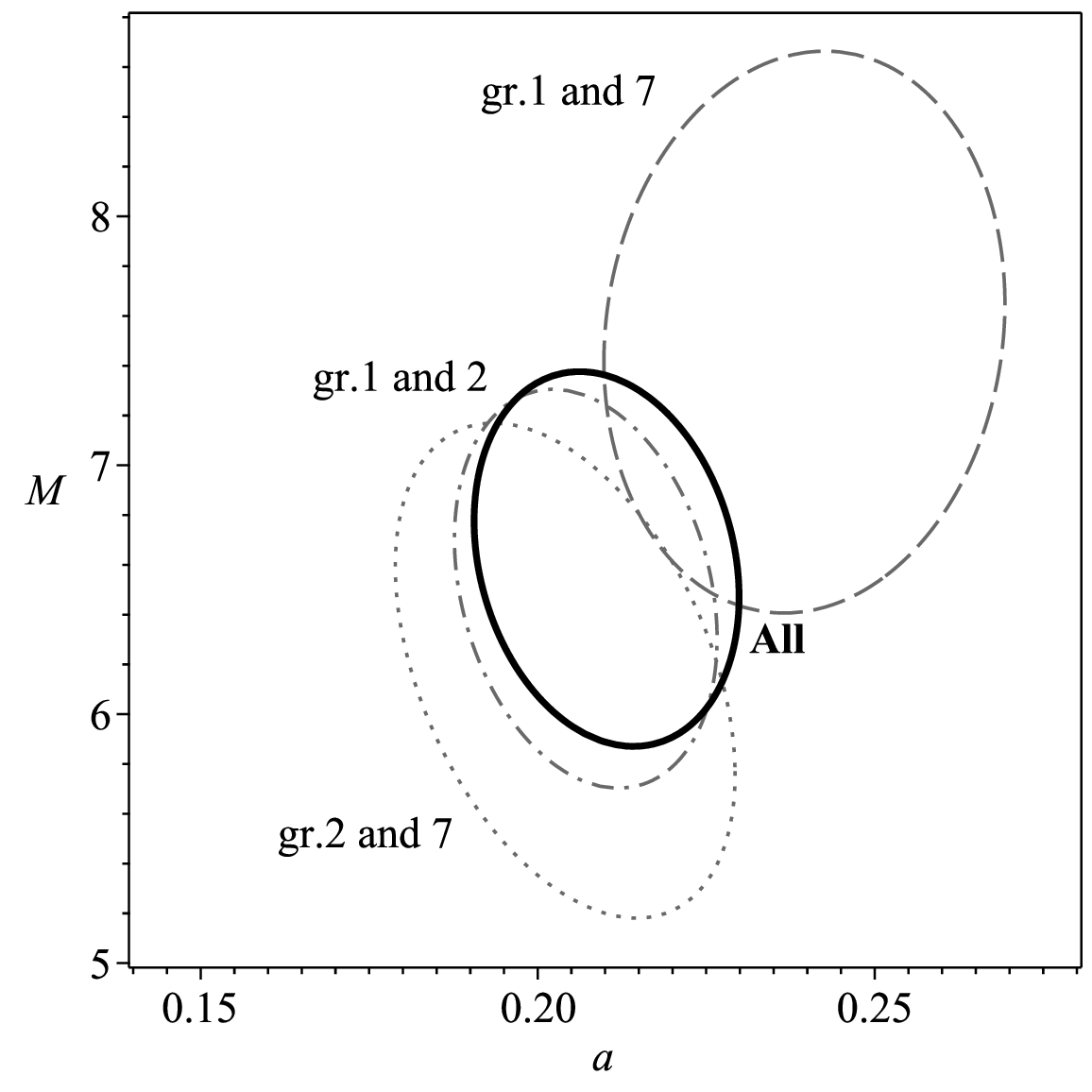, width=0.65\textwidth}
				\caption{The  1-$\sigma$ covariance plot obtained by the combination of two and three triads of QPOs.}} \label{3triads}
		\end{center}
	\end{figure}
	
	The confidence regions obtained with groups 1 and 7 (dotted line) and groups 2 and 7 (dashed line) barely touch each-other. The dash-dotted line which represents confidence region resulting from the combination of groups 1 and 2 overlaps partially with both previous. All of the thin line ellipses overlap partially with the solid one which is a result of the combination of all three of the considered groups of observations.

	\section*{5. Discussion}\label{Discussion}
	The aim of the current paper is to apply Bambi's  method (Bambi, 2015) to a source which contains two or more simultaneous triads of variability components. 
	The joint $\chi^{2}$ variable that can be composed in this situation is an authentic one, in a sense that it has non-zero degrees of freedom, a fact which allows the goodness of the fit to be tested. For this purpose we chose the accreting millisecond X-ray pulsar IGR~J17511-3057, whose X-ray spectrum contains three supposedly simultaneous triads of variability components described in Section 2. 
	
	The following aspects of the model are tested: the metric and the association of the observed frequencies with orbital frequency, periastron precession frequency and the nodal precession frequency, as stated by the relativistic precession model. First, the space-time in the vicinity of the central object can be described by Kerr metric. This of course is questionable for neutron stars. However, as Stuchl\'{\i}k \& Kolo\v{s}, (2015) show the Kerr metric is applicable for the calculation of the epicyclic frequencies of neutron stars with moderate spin, $a<0.4$. This constraint is satisfied by the individual predictions for the spin of the studied object obtained with groups 1 and 2 but not with group 7. All of the constraints obtained by the joint $\chi^{2}$ of two or three groups satisfy the cited theoretical constraint on the spin in question. 
	
	Second, it is customary to describe, according to the RPM, the HF QPOs of black holes and the kHz QPOs of neutrons stars by the orbital frequency and periastron precession frequency. It is not so common, however, to apply the nodal precession frequency to the hump component, instead of the usual C-type LF QPOs . An example for the latter can be seen in (van Straaten et al., 2003).     
	
	The limitations on the experimental side of this study are related mainly with the choice of the source. Indeed the presence of three triads of simultaneous variability components in the X-ray PDS of IGR~J17511-3057 which can be modeled by the relativistic precession model makes it an excellent choice. However, as the  Kalamkar et al. (2011) state and as we reiterate here in Section 2 the identification of the features in the X-ray PDS is not certain. Other sources that could serve to test the feasibility of the RPM through the methodology proposed in this study are: 4U 1728-34, XTE J1807-294 and 4U 1608-52, to name a few.    
	
	\begin{figure}[ht!]
		\begin{center}
			\centering{\epsfig{file=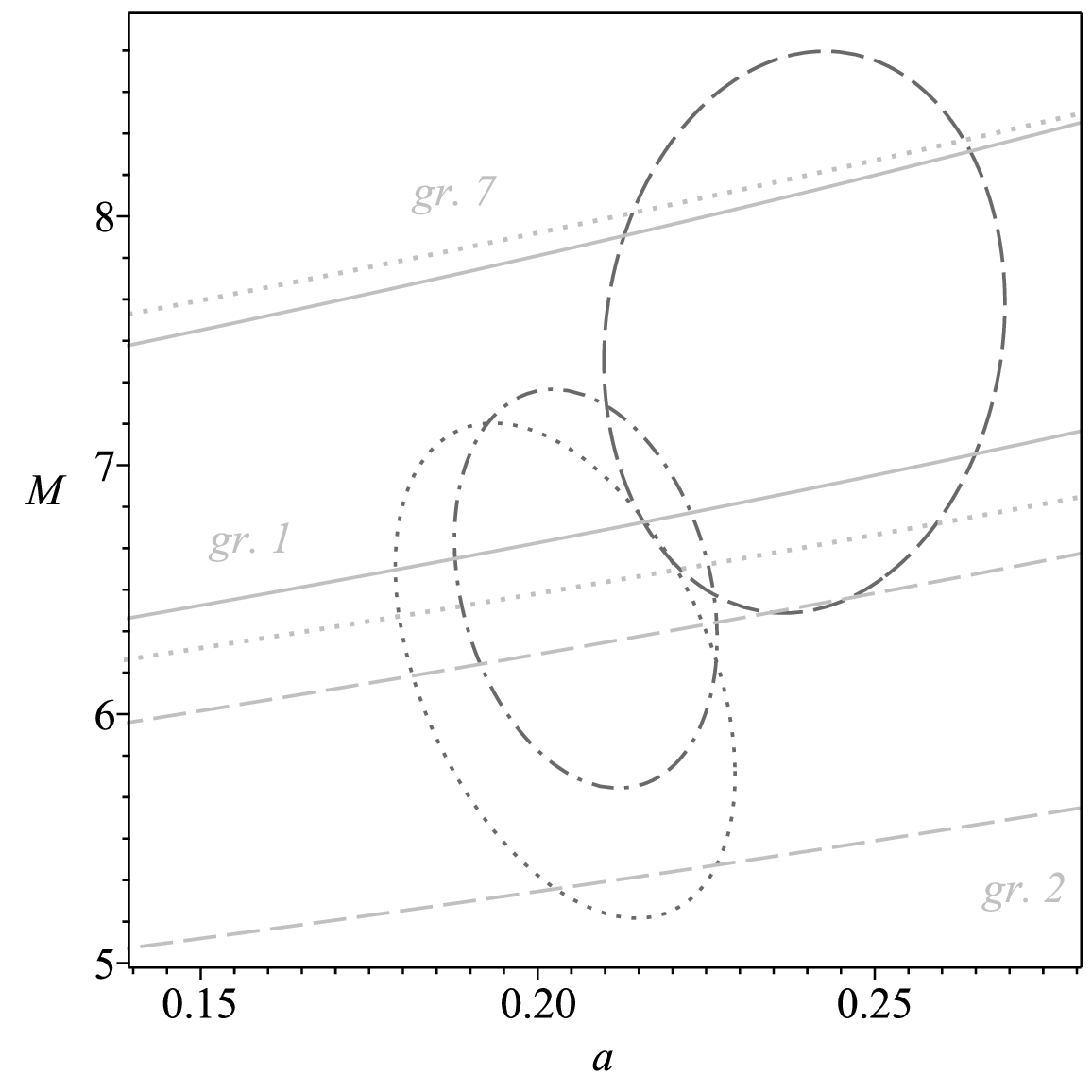, width=0.65\textwidth}
				\caption{The predictions for the $a-M$ relation coming from groups 1 and 2  from (Stefanov, 2016.}}\label{Stripes}
		\end{center}
	\end{figure}	
	The results that we obtain from the current study are as follows. The individual predictions of groups 1 and 2 for the mass and spin of the object are in marginal agreement -- the confidence regions obtained with them barely touch each other at the chosen level of confidence of 68.3\%. The prediction of group 7 is in conflict with them. The individual and the joint predictions for the spin coming from groups 1 and 2 are moderate and do not give us reason to question the applicability of Kerr metric. Group 7 alone predicts spin that is way too high to justify the use of Kerr metric.   
	
	The joint $\chi^{2}$ of groups 1 and 2 takes acceptable minimum value $\chi^2_{\rm min}$, i.e. these two observations can be reconciled. The inclusion of group 7 in the combinations of two or three groups deteriorates the goodness of the fit. In these cases the minimum values of the joint $\chi^{2}$ is much grater than the critical value for the chosen confidence level of 5 \% and favors the rejection of the model. It appears that group 7 is an outlier.
	
	With all versions of the merit function that we use here the estimates of the mass are much higher than the upper theoretical bound for a neutron star. Similar overestimation of the mass has been reported also for the accreting millisenod pulsar XTE J1807-294 (Tasheva \& Stefanov, 2019; Tasheva \& Stefanov, 2021). This drawback of the RPM occurs also with Z sources but not with atoll sources. See (Lin et al., 2011) and (Stefanov, 2020a).
	
	How do the results from this study compare with those of the earlier ones (Stefanov, 2015), (Stefanov, 2016) and (Tasheva, 2018)? The two latter studies provide constraints only for the mass, not for the spin because $\nu_{\rm h}$ was not taken into account. The degeneracy of the spin-mass relation which occurs when only two of the frequencies in the triad are use is reported also by (T\"{o}r\"{o}k et al., 2010; T\"{o}r\"{o}k et al., 2011; T\"{o}r\"{o}k et al., 2012 and Stefanov, 2016) and studied in more details in (Stefanov, 2020b). In the two studies different methodology for the estimation of the uncertainty of the mass was used which results in small differences in the predictions for the masses. Stefanov, (2016) states -- for given value of the spin the predictions of groups 1 and 2 for the mass  do not agree with each-other. This result is reproduced here in Figure 3. The stripes which represent the predictions coming from group 1, with solid gray boundaries and from group 2 -- dashed gray boundaries, do not overlap. They could be reconciled, however, if the uncertainty in the predictions for the spin was taken into account.
	
	In (Stefanov, 2016) agreement between groups 1 and 7 was ascertained. The latter agreement is questionable since it appear only for very high spins, a situation for which Kerr metric is not justified (Stuchl\'{\i}k \& Kolo\v{s}, 2015). This agreement disappears when the third frequency is included and the degeneracy is broken as a result. 
	
	On Figure 3 the ellipses coming from the combinations of two groups of observations from the present study are also given. It can be clearly seen that the joint prediction of groups 1 and 2, given by the dash-dotted ellipse overlaps with both the group 1 and group 2 stripes.
	
	In (Stefanov, 2015) and (Tasheva, 2018) the $\nu_{\rm u} - \nu_{\rm l}$  correlation is modeled and $\chi^{2}$ minimization is used for the obtaining of constraints on the mass and spin of the neutron star IGR~J17511-3057. In these papers only the pairs of the kHz QPOs present in power-density spectrum, i.e. the low-frequency features are not taken into account. The aim of (Tasheva, 2018) is to compare several alternative models for the kHz QPOs. As the author states all of the studied models  fail to fit the ensemble behavior of the lower and the upper frequency QPOs, according to the $\chi^{2}$ test. All but one of the considered models predict too high masses. 
	
	A solutions of the problem with the overestimation of could come from alternative gravity  theories in which, for example, a fundamental scalar field is present. We refer the reader to (Staykov et al., 2019; Marinov, 2020; Kuan et al., 2021; Staykov et al., 2023) and references therein for recent results on neutron stars coupled to a fundamental scalar fields. Some cosmological implications of massless and massive fundamental scalar fields can be found in (Arik \& Susam, 2021) and (Aditya \& Divya Prasanthi, 2023), respectively.
	
	Another idea for future study is to define a joint $\chi^{2}$ variable which contains frequencies from different incomplete simultaneous triads. A method which could be applicable to situations of insufficient data. 
	\section*{6. Conclusion}\label{Conclusion}
	
	If we assume that one or all of the frequencies in group 7 are outliers and remove this group from our considerations then minimum value of the joint $\chi^{2}$ variable that we obtain is acceptable and does not give us reason to question the feasibility of the model. 
	
	The constraints for the spin of the central neutron star in IGR~J17511-3057 obtained with groups 1 and 2, both the individual and the joint, are moderate and justify the use of the Kerr metric.
	
	The major deficiency of the model is, as it appears, the prediction for the mass of the neutron star. The constraints that we obtain in  all of the cases, including with the more favorable groups of observations 1 and 2, are much higher than the upper bound for a neutron star of three Solar masses. 
	\section*{Acknowledgments}
	I.S. would like to thank his wife for the support, Dr. Sava Donkov and Dr. Radostina Tasheva for the numerous discussions on the subject and prof. Stoytcho Yazadjiev for drawing his attention to the subject of QPOs.

	\section*{Appendix}\label{appendix}
	Explicit formulas for the orbital frequency $\nu_{\rm \phi}$ and the two epicyclic frequencies -- the radial $\nu_r$ and the vertical $\nu_{\theta}$ (Aliev \& Gal'tsov 1981; Aliev et al., 1986; Aliev et al., 2013)
	\begin{equation}
		\nu_{\rm \phi} =\left({1\over 2\pi}\right)\frac{ M^{1/2}}{ r^{3/2} + a M^{1/2}}\,\,,
		\label{orbf}
	\end{equation}
	\begin{equation}
		\nu_{r}^2 = \nu_{\rm \phi}^2\, \left( 1-\frac{6 M}{r} -\frac{3
			a^2}{r^2} + \, 8 a {M^{1/2}\over r^{3/2}}
		\right),
		\label{kerrradf}
	\end{equation}
	\begin{equation}
		\nu_{\theta}^2= \nu_{\rm \phi}^2\, \left(1
		+\frac{3 a^2}{r^2} - \, 4 a {M^{1/2}\over r^{3/2}} \right),
		\label{kerraxif}
	\end{equation}
	valid for the Kerr black hole. A change in the orientation of the orbit (direction of rotation of the hot spot) is equivalent to a change of the direction of rotation of the central object, i.e. a change in the sign of $a$.
\end{document}